\documentclass[fleqn,usenatbib]{mnras}

\usepackage{newtxtext,newtxmath}
\usepackage[T1]{fontenc}

\DeclareRobustCommand{\VAN}[3]{#2}
\let\VANthebibliography\thebibliography
\def\thebibliography{\DeclareRobustCommand{\VAN}[3]{##3}\VANthebibliography}

\usepackage{graphicx}	% Including figure files
\usepackage{amsmath}	% Advanced maths commands
\usepackage{caption} 
\usepackage{subfig}
\usepackage{subfloat}
\usepackage{float}
\usepackage{changes}
\usepackage{subcaption}
\usepackage{soul}
\usepackage{xcolor}
\usepackage[normalem]{ulem}

\usepackage{ulem}

\title[MSC as Sources of HE Gamma Radiation]{Massive Star Clusters as sources of high-energy gamma radiation}

\author[L. N. Padilha et al.]{
Luana N. Padilha,$^{1}$\thanks{E-mail: luana.natalie.padilha@uel.br}
Rita C. Anjos,$^{1,2,3,4,5,6}$
\\
$^{1}$Departamento de Física, Universidade Estadual de Londrina (UEL), Rodovia Celso Garcia Cid Km 380, 86057-970 Londrina, PR, Brazil\\
$^{2}$Departamento de Engenharias e Exatas, UFPR, Pioneiro, 2153, 85950-000 Palotina, PR, Brazil\\
$^{3}$Max-Planck-Institut für Kernphysik, Saupfercheckweg 1, D-69117 Heidelberg, Germany\\
$^{4}$ Programa de Pós-Graduação em Física e Astronomia, UTFPR, Av. Sete de Setembro, 3165, 80230-901, Curitiba, PR, Brazil\\
$^{5}$Programa de Pós-Graduação em Física Aplicada, Unila, 85867-670, Foz do Igua\c{c}u, PR, Brazil\\
$^{6}$N\'ucleo de Astrof\'{\i}sica e Cosmologia (Cosmo-Ufes), Universidade Federal do Esp\'irito Santo, 29075--910, Vit\'oria, ES, Brazil
}

\date{Accepted XXX. Received YYY; in original form ZZZ}

\pubyear{\the\year{}}

\begin{document}
\label{firstpage}
\pagerange{\pageref{firstpage}--\pageref{lastpage}}
\maketitle

% Abstract of the paper
\begin{abstract}
This paper investigates the contribution of massive star clusters (MSC) as sources of high-energy gamma rays and their impact on the ultra-high-energy (UHE) emission observed throughout the Galaxy. By modeling proton injection, the study explores how the acceleration of protons in massive star clusters contributes to the gamma radiation detectable from Earth. The analysis focuses on two primary types of clusters: widespread, dispersed clusters and younger, compact massive clusters, both of which host shock waves generated by supernova remnants (SNR). Clusters located near the solar system, within a 3-kiloparsec radius, are identified. Analytical methods are used to calculate energy spectra and gamma-ray production rates. The findings suggest that young and compact MSC contribute to multi-TeV to PeV gamma-ray emission, with the dominant contribution arising from nearby populations.
\end{abstract}

\begin{keywords}
gamma-rays: stars - open clusters and associations: general - cosmic rays - astroparticle physics - acceleration of particles
\end{keywords}

%%%%%%%%%%%%%%%%%%%%%%%%%%%%%%%%%%%%%%%%%%%%%%%%%%

%%%%%%%%%%%%%%%%% BODY OF PAPER %%%%%%%%%%%%%%%%%%

\section{Introduction}

Recent advances in measuring the composition of Cosmic Rays (CR) above the spectral ``knee" \footnote{A discontinuity in the CR spectrum occurs around $10^{15}$ eV.} have significantly aided our understanding of their sources. Studies of atmospheric showers have revealed that lighter particles, such as protons and helium nuclei, dominate at energies around  $10^{17}$ eV,  just above the knee in the CR spectrum~\citep{2021EPJC...81..966A, 2016Natur.531...70B}. As particle energies increase beyond $10^{17}$ eV, the overall composition of CR tends to become heavier, although lighter nuclei can still be present~\citep{2022Galax..10...75S, 2018AdSpR..62.2764B}. Models have been proposed to explain particle acceleration at the knee, including galactic wind termination shocks~\citep{1987ApJ...312..170J}), stellar environments~\citep{1988ApJ...333L..65V}, and Supernovae (SNe) collapse in dense wind environments~\citep{2010ApJ...718...31P}. The theory that most CRs are accelerated by shocks from SNR has gained significant attention (for a review, see e.g.~\citet{1988ApJ...333L..65V, 2004A&A...424..747P, 2012JCAP...07..038C,2013MNRAS.431..415B, 2021MNRAS.504.6096M,2024PhRvD.110d3046K}). Gamma-ray observations provide strong evidence supporting CR acceleration in SNR. The observed increase in $\pi^0$ decay within SNR spectra confirms the presence of hadronic and leptonic processes~\citep{2013Sci...339..807A,2019A&A...623A..86A,2022hxga.book...52V}. Furthermore, the detection of gamma rays with energies in the TeV range in young SNR further strengthens this theory, although such high-energy emissions are less prominent in middle-aged SNR, where a spectral break occurs around 100 GeV~\citep{2006A&A...449..223A}. For SNR to act as sources of PeV cosmic rays, their parameters, particularly the shock velocity, would need to reach extreme values. Even under such conditions, the maximum energy achieved would still be insufficient to fully account for the highest energies observed in the cosmic ray spectrum~\citep{2023MNRAS.519..136V}.

\begin{figure*}
  \centering
  \includegraphics[angle=0,width=\textwidth]{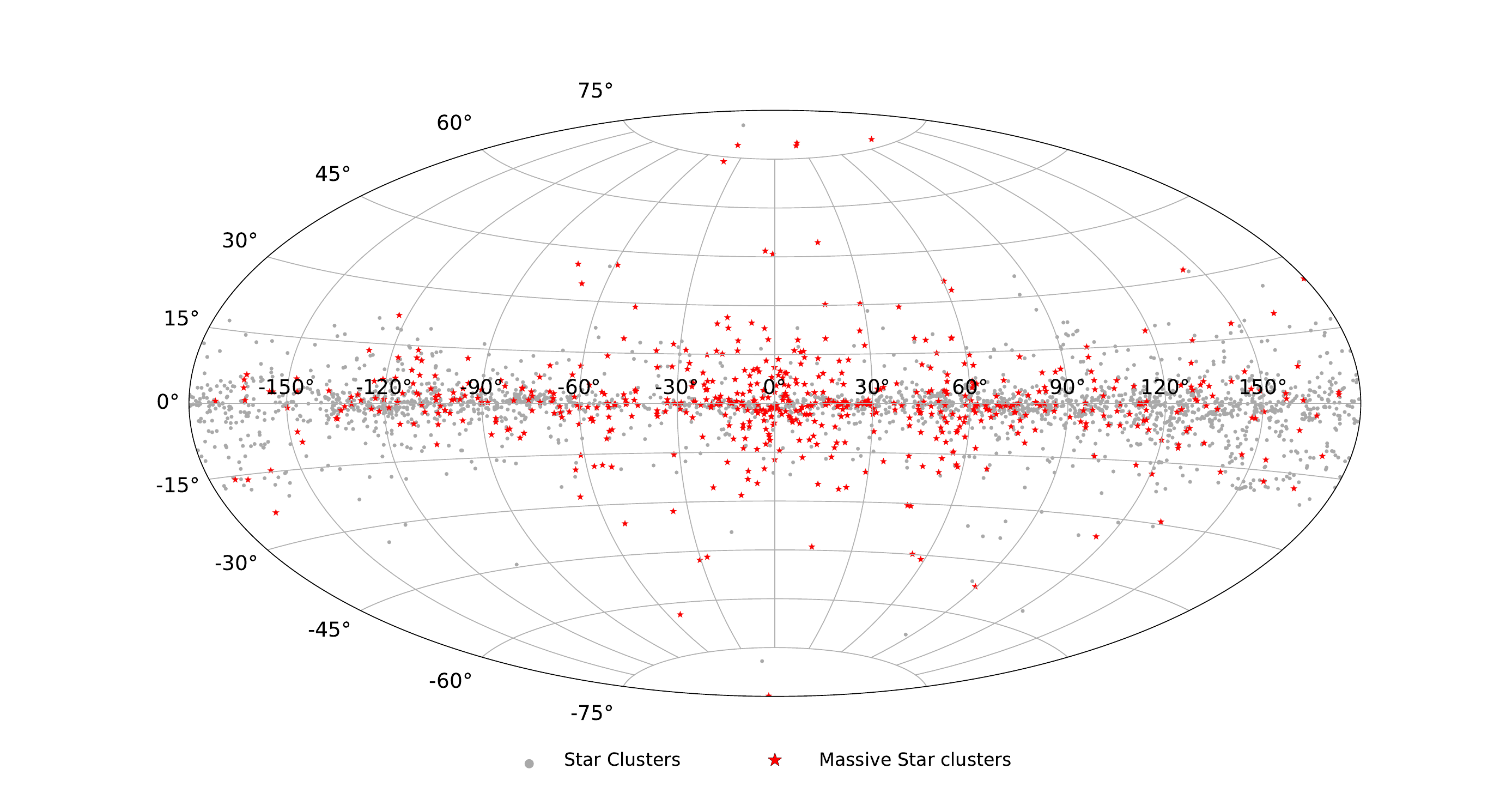}
  \caption{Mollweide projection of stars from the~\citet{2013A&A...558A..53K} catalog, plotted in Galactic coordinates. Star clusters are shown in gray, while massive star clusters are highlighted in red.}
  \label{fig:skymap}
\end{figure*}

Iron abundances suggest that CRs may be accelerated in OB star clusters and superbubbles \citep{2008NewAR..52..427B,2016ApJS..223...26A}. Some models incorporate the combined effects of SNe and stellar winds from young massive stars within superbubbles \citep{2001AstL...27..625B,2004A&A...424..747P,2010A&A...510A.101F}. Massive star clusters (MSC) or superclusters, with masses of several thousand solar masses, can host thousands of young stars in compact cores~\citep{2010ARA&A..48..431P,2018AdSpR..62.2764B}. These clusters are recognized as powerful particle accelerators within our Galaxy, producing strong stellar winds that collide and create shock waves capable of accelerating particles to high energies~\citep{2021MNRAS.504.6096M,2023MNRAS.519..136V}. Evidence of CR acceleration has been detected by Fermi-LAT in compact clusters such as Cygnus OB2~\citep{2011Sci...334.1103A}, Westerlund 2~\citep{2018A&A...611A..77Y} and Westerlund 1~\citep{2012A&A...537A.114A, 2013MNRAS.434.2289O, 2019NatAs...3..561A}. Furthermore, gamma-ray emission have been analyzed from the NGC 3603 cluster~\citep{2017A&A...600A.107Y}, the Cygnus Cocoon~\citep{2022AdSpR..70.2685B}, and HESS J1646-458, which is also associated with Westerlund 1~\citep{2012A&A...537A.114A}. These studies indicate that gamma rays are produced through interactions of CRs with the surrounding gas, primarily through the production and decay of pions~\citep{2019RLSFN..30S.159Y,2021JCAP...10..023D,coelho2022updated}. These clusters exhibit strong potential for accelerating particles to energies reaching up to PeV.  

In this paper, we investigate how proton acceleration in MSC contributes to high-energy gamma-ray emission. Our analysis focuses on two primary types of clusters: dispersed clusters and younger, compact massive clusters, both of which host shock waves generated by SNR. For our analysis, we used clusters identified near the solar system. The paper is organized as follows. Section~\ref{sec:2} presents an overview of the Galactic star cluster population considered in this study, including an analysis of the observational catalog, estimates of the corresponding supernova rate, and the extension of the sample through a synthetic population to account for incompleteness. Section \ref{sec:3} presents the particle spectrum generated by our model, based on the methods developed by \citet{2023MNRAS.519..136V}. In Section \ref{sec:4}, we discuss gamma-ray emission in detail. Finally, we present the key findings and general insights derived from this work in Section \ref{sec:5}.

%mudei o titulo 
\section{Cluster Population analysis based on Observational and Synthetic Data}\label{sec:2}

\begin{figure*}
   \centering
    \subfloat[MSC survey clusters]{\includegraphics[angle=0,width=0.48\textwidth]{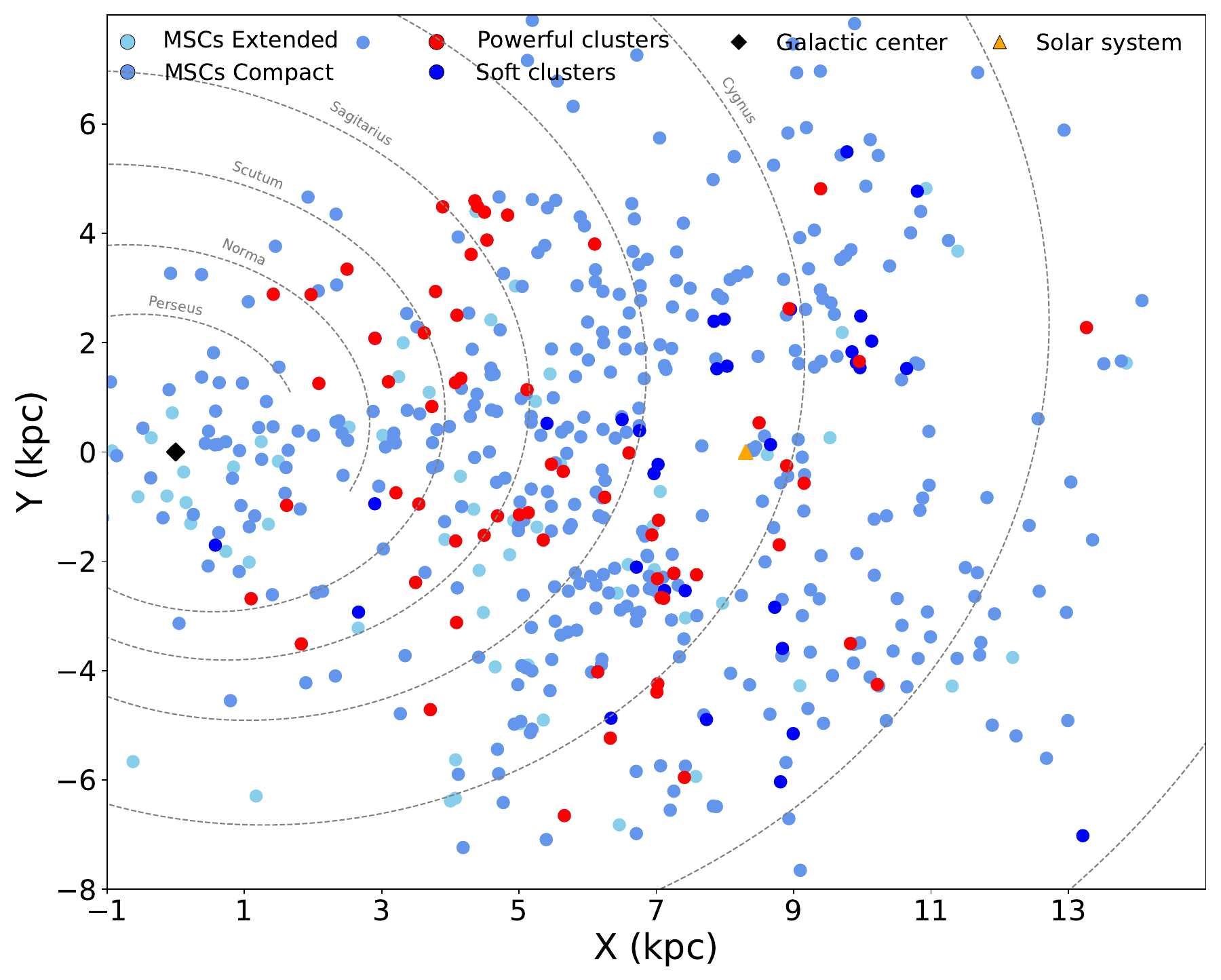}}
    \subfloat[Zoom into MSC clusters within a $3\, \mathrm{kpc}$ radius]{\includegraphics[angle=0,width=0.48\textwidth]{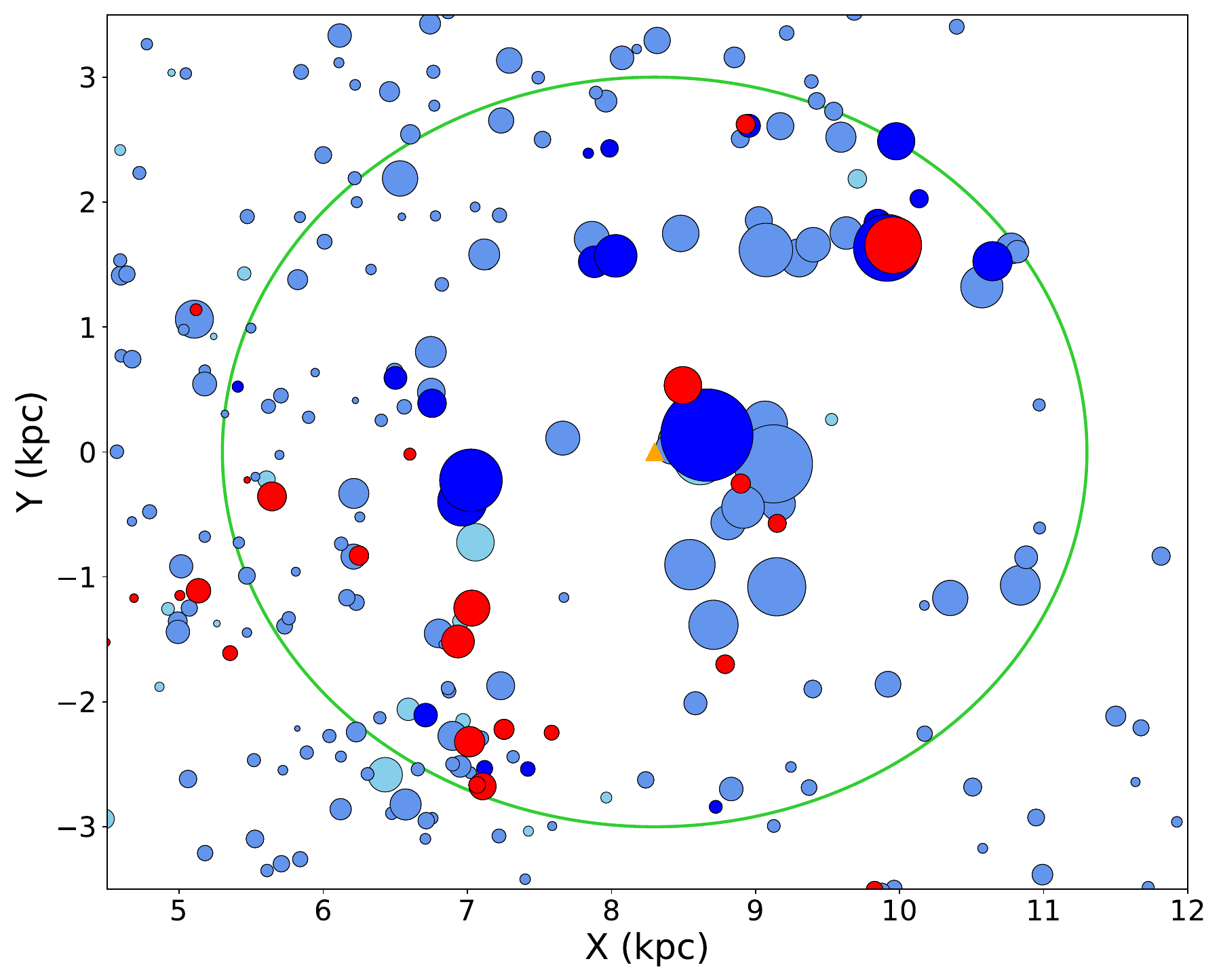}}  
    \caption{Distribution of star clusters from the MSC survey \citep{2013A&A...558A..53K} is shown projected onto the Galactic XY plane, with distances measured relative to the Sun. The right panel displays a zoomed-in view of nearby clusters, where circle sizes scale with stellar membership. A green circle highlights sources within $3\, \mathrm{kpc}$ of the Solar System.}
    \label{fig:xymap}
\end{figure*} 
  
Stellar cluster catalogs provide essential information for understanding galactic structure evolution and star formation processes. Through multi-wavelength photometric observations, they enable precise determination of cluster properties, including age, mass, metallicity, and spatial distribution across our Galaxy~\citep{2008A&A...477..165P, 2012A&A...543A.156K, 2013A&A...558A..53K, 2018sas..conf..222K, 2018A&A...614A..22P, 2019INASR...3..324P, 2019INASR...3..330K, 2020A&A...640A...1C, 2023A&A...672A.187J, 2025A&A...697A.129K}.

In this paper, we use the catalog presented by~\citet{2013A&A...558A..53K}\footnote{\href{https://vizier.cds.unistra.fr/viz-bin/VizieR?-source=J/A\%2BA/558/A53}{Accessible at J/A+A/558/A53}}. It is based on data from the PPMXL catalog~\citep{2010AJ....139.2440R} and the 2MASS catalog~\citep{2006AJ....131.1163S}. The PPMXL provides information on approximately 900 million objects, including coordinates, proper motions, and low-precision photometry, while 2MASS offers precise magnitudes in three bands (J, H, Ks) for about 400 million entries~\citep{2006AJ....131.1163S}. The authors also used a combined version of these catalogs, known as 2MAst, to verify their list of input clusters and assist in determining astrometric and photometric parameters. 

For our study of star clusters as CR and gamma-ray sources, we selected the~\citet{2013A&A...558A..53K} catalog due to its extensive sky coverage and wealth of observational parameters. The catalog identifies 3.006 clusters comprising approximately 400.000 probable member stars (kinematic and photometric association probability > 60\%), with an average of 100 stars per cluster. We further refined this sample to study clusters with specific properties relevant for particle injection and subsequent gamma-ray emission analysis. Our study focuses specifically on MSC, which are particularly efficient particle accelerators due to their population of young, massive stars (>8 $\mathrm{M_{\odot}}$~\citep{2008ApJ...675..614P}). The extreme pressure conditions in these stellar cores lead to rapid nuclear fuel consumption, limiting their lifetimes to just a few million years~\citep{2017imas.book.....C}. Figure~\ref{fig:skymap} shows the Galactic distribution of massive and lower-mass star clusters included in our sample.

Star clusters are known to be numerous throughout the Galaxy and represent favorable environments for particle acceleration to high energies. However, only the massive clusters have enough power to inject a non-negligible amount of particles into the interstellar medium. To identify these candidates, we implemented a multi-step filtering process on the initial catalog of 3.006 clusters. Our first selection criterion required complete parameter sets for all clusters, reducing the sample to 2.961 objects. To identify massive clusters capable of efficient particle acceleration, we employed the tidal radius as a key diagnostic parameter. The tidal radius, which marks the boundary where Galactic tidal forces overcome a cluster's self-gravity \citep{2013ApJ...764..124W}, constitutes a mass indicator for stellar clusters. Clusters with larger tidal radii generally have stronger gravitational potentials, making them more likely to retain accelerated particles and contribute to the interstellar particle population.

Assuming that the cluster’s gravitational field corresponds to a circular orbit within a spherically symmetric galactic potential, the tidal radius can be approximated. According to first-order tidal theory, the tidal radius is defined by~\citet{1957ApJ...125..451V} and~\citet{2013ApJ...764..124W}:
\begin{equation}     
r_t = R_{\mathrm{gc}}\left( \frac{M_{\mathrm{cl}}}{M_{\mathrm{gal}}} \right)^{1/3}
\label{eq:mass}
\end{equation}
where $R_{\mathrm{gc}}$ is the galactocentric distance of the cluster, $M_{\rm cl}$ is the mass of the cluster, and $M_{\rm gal}$ is the mass of the Galaxy. Since our catalog is incomplete and contains spatial gaps, we adopted a characteristic Galactic mass of $10^{11}\,\mathrm{M_{\odot}}$ for systems within 100\,kpc \citep{1998MNRAS.294..429D}. A comparison between our cluster mass estimates and those obtained by other methods in the literature \citet{2018MNRAS.478.1520B} shows that our masses are systematically lower on average. Applying Eq.~\ref{eq:mass}, we derived individual cluster masses, classifying those above $10^{3}\,\mathrm{M_{\odot}}$ as massive. This selection yielded 598 massive clusters, representing $20.19\%$ of the catalog sample.

For SNe to occur within star clusters, the presence of massive stars is essential, as these undergo core-collapse within $<40\,\mathrm{Myr}$~\citep{2008ApJ...675..614P}. Applying this age cut to our MSC sample yields 98 clusters ($\sim\!16.36\%$ of MSC), representing potential SN progenitors. The most compact MSC provide particularly efficient particle acceleration environments. We characterize cluster density using the King radius $r_0$ (where stellar density drops to $\sim\!1/3$ of its central value~\citep{1962AJ.....67..471K}, adopting $r_0 \lesssim 5\,\mathrm{pc}$ as in~\citet{2023MNRAS.519..136V}. For age classification, we reconcile literature thresholds ranging from $\lesssim\!10\,\mathrm{Myr}$~\citep{2023MNRAS.519..136V,2025A&A...695A.175M} to $\lesssim\!30\,\mathrm{Myr}$~\citep{2024A&A...686A.118C,2024ApJ...972L..22P} by adopting an intermediate $ \lesssim 20\,\mathrm{Myr}$ criterion for `young` clusters, balancing these approaches while maintaining physical relevance to massive star evolution timescales.

Young massive clusters in our sample exhibit enhanced stellar winds compared to older populations, as they contain stars that have not yet undergone SNe. We classify these systems into two categories: \textit{Powerful clusters}: young ($\lesssim 20\,\mathrm{Myr}$), compact ($r_0 \lesssim 5\,\mathrm{pc}$), and massive ($>10^3\,\mathrm{M_\odot}$) systems with intense collective winds, and \textit{Soft clusters}: the remaining massive clusters ($20-40\,\mathrm{Myr}$) spanning both compact and extended configurations. The wind termination shock, a key particle acceleration region, forms where these collective stellar winds interact with the interstellar medium (ISM), creating a shock boundary that efficiently energizes particles. Powerful clusters dominate this process due to their exceptional wind luminosity. These clusters also exhibit high stellar densities (\(\gtrsim 10^2\,\mathrm{stars\,pc^{-3}}\)).
When combined with subsequent SNe events, both cluster types contribute to enriched environments through different mechanisms: Powerful clusters primarily through wind shocks, while Soft clusters exhibit acceleration processes powered by SNe because their collective winds are less energetic due to the older age of their stars. Our model incorporates these distinct emission regimes, capturing their relative contributions to cosmic ray production.

Figure~\ref{fig:xymap}-(a) displays our filtered cluster sample in Galactic Cartesian coordinates. To study local contributions to cosmic-ray populations, we implemented a 3 kpc spatial cutoff, selecting only the nearest clusters, see Figure~\ref{fig:xymap}-(b). Within this volume, our classification yields 15\% Soft clusters and 11\% Powerful clusters, reflecting their relative abundance in the solar neighborhood.

\subsection{Simulated massive star cluster population}\label{sec:2.2}

Observational constraints, including sensitivity limits and interstellar extinction, hinder the detection of faint or distant star clusters in the Milky Way. As a result, existing catalogs are inherently incomplete, preventing accurate determinations of the total cluster population and their Galactic distribution~\citep{2025A&A...695A.175M}.

\begin{figure}
    \centering
    \includegraphics[width=1.0\linewidth]{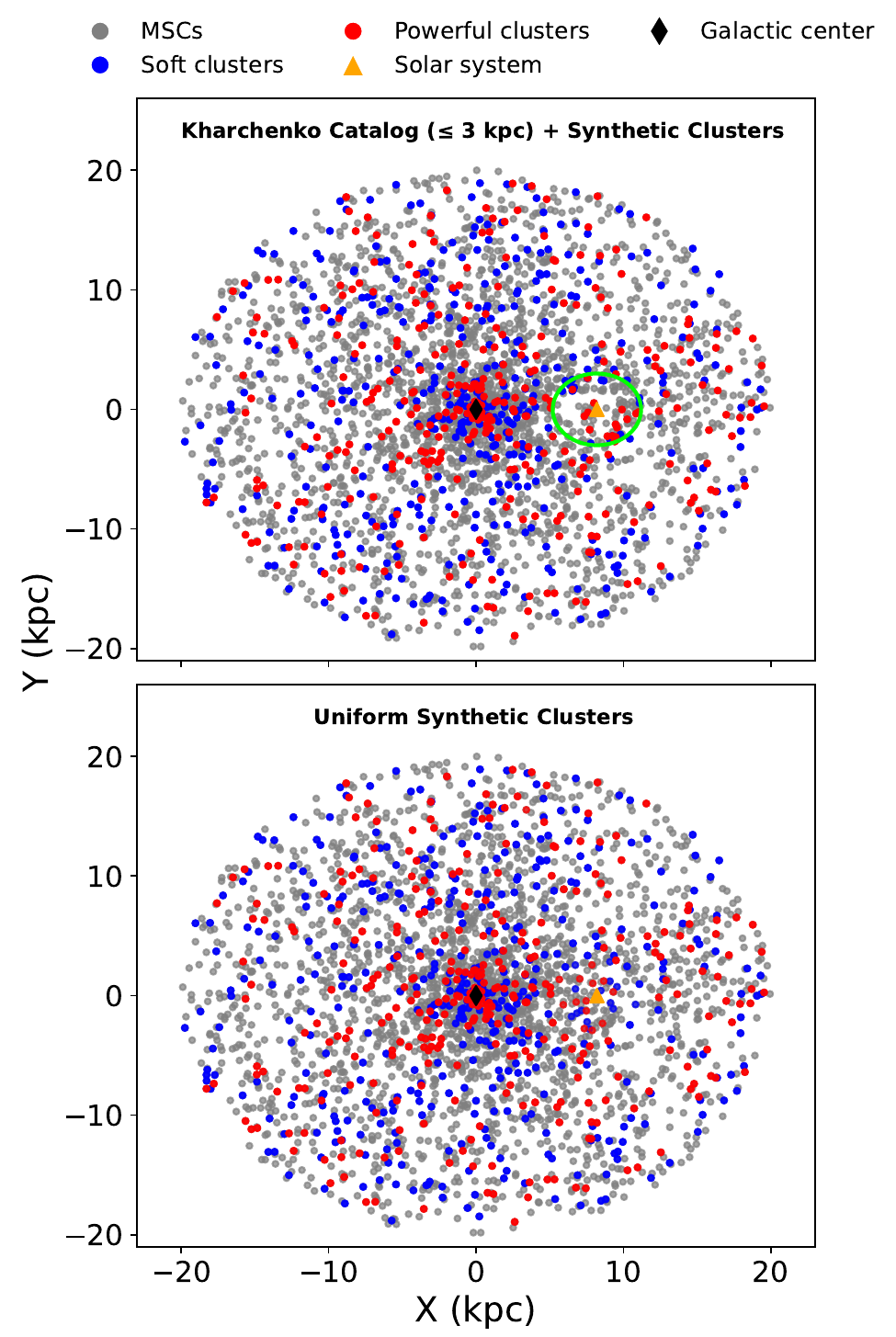}
    \caption{Spatial distribution of MSC in the Milky Way, obtained by combining a synthetic population with observational data and projected onto the Galactic XY plane. The green circle denotes sources within $3.0\,\mathrm{kpc}$ of the Solar System, as listed in the catalog by \protect\citet{2013A&A...558A..53K}.}
    \label{fig:mapsim}
\end{figure}

Star cluster formation in the Milky Way occurs at a rate of $0.2-0.5\,\mathrm{Myr^{-1}\,kpc^{-2}}$, equivalent to $\sim 200\,\mathrm{M_{\odot}\,Myr^{-1}\,kpc^{-2}}$ according to \citet{2006A&A...445..545P} and \citet{1991MNRAS.249...76B}. When extrapolated over a minimum cluster lifetime of $250\,\mathrm{Myr}$, this suggests the Galaxy should host between $23.000$ and $37.000$ star clusters \citep{2010ARA&A..48..431P}. However, current observational catalogs reveal a significant discrepancy, with \citet{2020A&A...640A...1C} reporting only $\sim 2000$ clusters and \citet{2013A&A...558A..53K} listing $3006$ clusters. This implies that only $10\%-15\%$ of the expected population has been identified, underscoring the significant gaps in existing surveys.

To address this limitation, our analysis uses the \citet{2013A&A...558A..53K} catalog, which is considered complete for clusters within $2\,\mathrm{kpc}$ of the Sun. While undetected clusters likely exist beyond $3\,\mathrm{kpc}$, our focus on young clusters ($<40\,\mathrm{Myr}$) minimizes the impact of this incompleteness. Nevertheless, to account for the expected population at larger galactocentric distances ($3 - 20\,\mathrm{kpc}$), we have supplemented the observational data with a synthetic population of MSC distributed uniformly across the Galactic disk. In contrast to \citet{2025A&A...695A.175M}, who rely on a fully synthetic population model with detailed spiral‐arm geometry and time‑dependent cluster formation rates, our hybrid scheme is both sufficient and effective: by anchoring to the observationally complete \citet{2013A&A...558A..53K} catalog within $3\,\mathrm{kpc}$, we ensure that the dominant gamma‑ray contributors are modeled with real data, while the uniformly distributed synthetic component beyond this radius introduces only minimal additional uncertainty. This combined approach provides a more realistic estimate of the Milky Way's MSC population while taking into account the limitations of current observations.
 
The synthetic population was constructed based on the estimate that the Galaxy hosts approximately 400,000 massive stars, with about 80\% residing in star clusters~\citep{2023MNRAS.519..136V}. 
To estimate the number of massive stars per cluster, we calculated the average using the mass-to-star number relation derived from clusters in the vicinity of the Solar System, based on the catalog by~\citet{2013A&A...558A..53K}. 
This yielded a mean of 100 massive stars per cluster, leading to an estimated total of approximately 3,200 MSC in the Milky Way. 
This estimate is consistent with previous studies, such as~\citet{2022MNRAS.515.2256V}, which report comparable numbers. While the catalog remains reliable out to $3\,\mathrm{kpc}$ (and fully complete to $2\,\mathrm{kpc}$), we use real data for nearby clusters ($<3\,\mathrm{kpc}$) and synthetic clusters beyond this limit, taking advantage of the fact that gamma-ray flux contributions are dominated by nearby sources to minimize potential biases from synthetic extrapolations.

For the Monte Carlo simulation, we first generated sources distributed across $0-20\,\mathrm{kpc}$, then selected those beyond $3\,\mathrm{kpc}$ while applying a spatial filter to prioritize clusters near the Galactic plane. We applied a spatial filter to focus on regions where massive star formation is most active. Since most young and massive clusters are found near the Galactic plane, we adopted a latitude cut of \(\left| b \right| < 5^{\circ }\) and a vertical height limit of  \( \left| z \right| \le  0.5\) kpc. The synthetic clusters were classified into soft and powerful types using the proportions derived in Section~\ref{sec:2} from observed clusters within $3\,\mathrm{kpc}$, resulting in $323$ powerful and $444$ soft clusters beyond this distance. This combined catalog of observed and synthetic clusters, see Figure~\ref{fig:mapsim}, was then used to model the gamma-ray flux. This approach ensures our analysis benefits from both observational data where available and statistically robust synthetic populations where observations are incomplete, providing a comprehensive view of MSC contributions to Galactic gamma-ray emissions.

\subsection{Supernova rates in Stellar Clusters}\label{sec:2.3}

To evaluate the efficiency of particle acceleration in MSC hosting SN explosions, it is first necessary to estimate the minimum total number of supernovae (\(N_{\mathrm{SN}}\)) occurring in these environments.  This estimate is crucial for determining the parameter \(n_{\mathrm{sn}}\), which represents the fraction of supernovae occurring at the edge of the cluster core, where they can generate fast shocks within regions dominated by collective stellar winds (see \citet{2023MNRAS.519..136V}). These conditions are particularly favorable for enhancing the efficiency of particle acceleration.

Previous studies have estimated SN rates in clusters using various methods. \citet{2021MNRAS.506.4131W} derived SN counts from iron abundance spreads in Galactic globular clusters. Their approach first calculates the iron mass required to produce observed abundance spreads, then estimates the number of SNe needed to contribute this iron, and finally derives SN rates by normalizing the SN count by the cluster's initial mass. Their results indicate SN rates of $10^{-2}-10^{-5}$ and total SN counts of $10^{1}-10^{4}$ per cluster. An alternative approach estimates the SN count via:
\begin{equation}    
N_{\rm SN} = N_{\rm \star} \times f_{\rm SN},
\label{eq:snr}
\end{equation}
where $N_{\rm \star}$ is the number of stars in the cluster and $f_{\rm SN}$ is the fraction of stars that undergo SN explosions. This fraction depends on the Initial Mass Function (IMF), stellar evolution rates, and metallicity. Stellar evolution theory suggests that stars with masses $\geq 8\, \mathrm{M_\odot}$ are viable core-collapse SN progenitors \citep{2009MNRAS.395.1409S}, whereas those with masses above $\geq 20\, \mathrm{M_\odot}$ typically collapse directly into black holes. Accordingly, we adopt the mass range \(8\text{ -- }20\,\mathrm{M_\odot}\) as representative of SN progenitors in this study.

For our IMF calculations, we use the Chabrier function \citep{2003PASP..115..763C}, appropriate for young clusters with stellar masses $>1\, \mathrm{M_\odot}$:
\begin{equation}
\xi(m) \propto m^{-2.35},
\end{equation}
where $m$ is stellar mass and the exponent follows \citet{1955ApJ...121..161S}. Integrating this IMF over $8-20$ $\mathrm{M_\odot}$ and normalizing by the total number of stars with $m > 1\, \mathrm{M_\odot}$ (integrated from 1 to 150 $\mathrm{M_\odot}$) yields $f_{\rm SN} \approx 0.043$. Thus, $\sim$ 4.3\% of stars above $1\, \mathrm{M_\odot}$ fall within the SN progenitor mass range. Our $f_{\rm SN}$ value agrees with \citet{2021MNRAS.506.4131W}'s findings. Applying  $f_{\rm SN} = 0.043$ reveals that all powerful clusters contain at least two SNe. Because we assume that $n_{\rm sn} = 50\%$, the effective number of SN for each cluster is \(\ge\)1. Considering that our method for estimating cluster masses tends to underestimate the true values (see Sec. \ref{sec:2}), the corresponding number of supernovae should therefore be regarded as a lower limit.

\section{Particle spectrum}\label{sec:3}

Soft clusters are compact stellar groups without collective winds but can still accelerate particles up to  
\[
E_{\rm soft}^{\rm max} \sim 0.1\, Z\, B\, E_{\rm SN}^{0.33}\, \rho^{-0.33}\, \upsilon_{5}^{0.33},
\]\footnote{This expression is derived from the Sedov--Taylor solutions~\citep{2009MNRAS.396.2065C}, applied to the formulation of the maximum energy from~\citet{1983A&A...125..249L}} 
reaching PeV energies through supernova and stellar wind shocks~\citep{1983A&A...125..249L,2005JPhG...31R..95H}.  
Here, $B$ is the magnetic field strength, $E_{\rm SN}$ is the canonical energy released by a core-collapse supernova, $\rho$ is the ambient density, $\upsilon_5$ is the shock velocity in units of $5000~\mathrm{km~s^{-1}}$, and $Z$ is the atomic number.  
Assuming typical environmental values for clusters in the Galaxy ($B \sim 2~\mu\mathrm{G}$ and $\rho \sim 0.01~\mathrm{cm^{-3}}$), and noting that small variations in these parameters have a negligible effect, the maximum energy can be approximated as  
\[
E_{\rm soft}^{\rm max} \sim 0.1\, Z\, \upsilon_{5}^{0.33}.
\]
In contrast, powerful clusters are young, compact systems characterized by high stellar densities, collective stellar winds, and strong turbulence driven by supernova activity.  
Under these conditions, particles can be accelerated up to~\citep{2023MNRAS.519..136V} 
\[
E_{\mathrm{pow}}^{\mathrm{max}} \sim 4\,\mathrm{Z}\,\upsilon_5\, f_c(R_c, N_{*}, n_c, \eta_T)\ \mathrm{PeV},
\]
where \(f_c\) is a function that encapsulates the dependence on cluster properties, including the cluster radius (\(R_c\)), the number of massive stars (\(N_{*}\)), the gas density (\(n_c\)), the fraction of injected stellar mechanical power, and the level of magnetohydrodynamic turbulence (\(\eta_T\)). Since the shock velocity is the dominant parameter and its influence outweighs the variations associated with the quantities in \(f_c\), the expression can be approximated as  
\[
E_{\mathrm{pow}}^{\mathrm{max}} \sim 4\,\mathrm{Z}\,\upsilon_5,
\]
assuming typical cluster parameters and a turbulence level of \(\eta_T \sim 10\%\).
In our model, Bohm diffusion is assumed for particle acceleration in supernova shocks occurring within clusters that possess collective winds. This assumption is justified by the strong turbulence and magnetic field amplification expected in such environments, which keep particle transport close to the Bohm limit.
The maximum energy may be further increased by magnetic turbulence and field amplification in the shocked medium~\citep{2006ApJ...636..140Z}, as well as by reacceleration processes occurring within extended stellar wind bubbles~\citep{1987ApJ...312..170J}.

\begin{figure*}
   \centering
    \subfloat[Composition spectrum ($Z \leq 40$)]{\includegraphics[angle=0,width=0.50\textwidth]{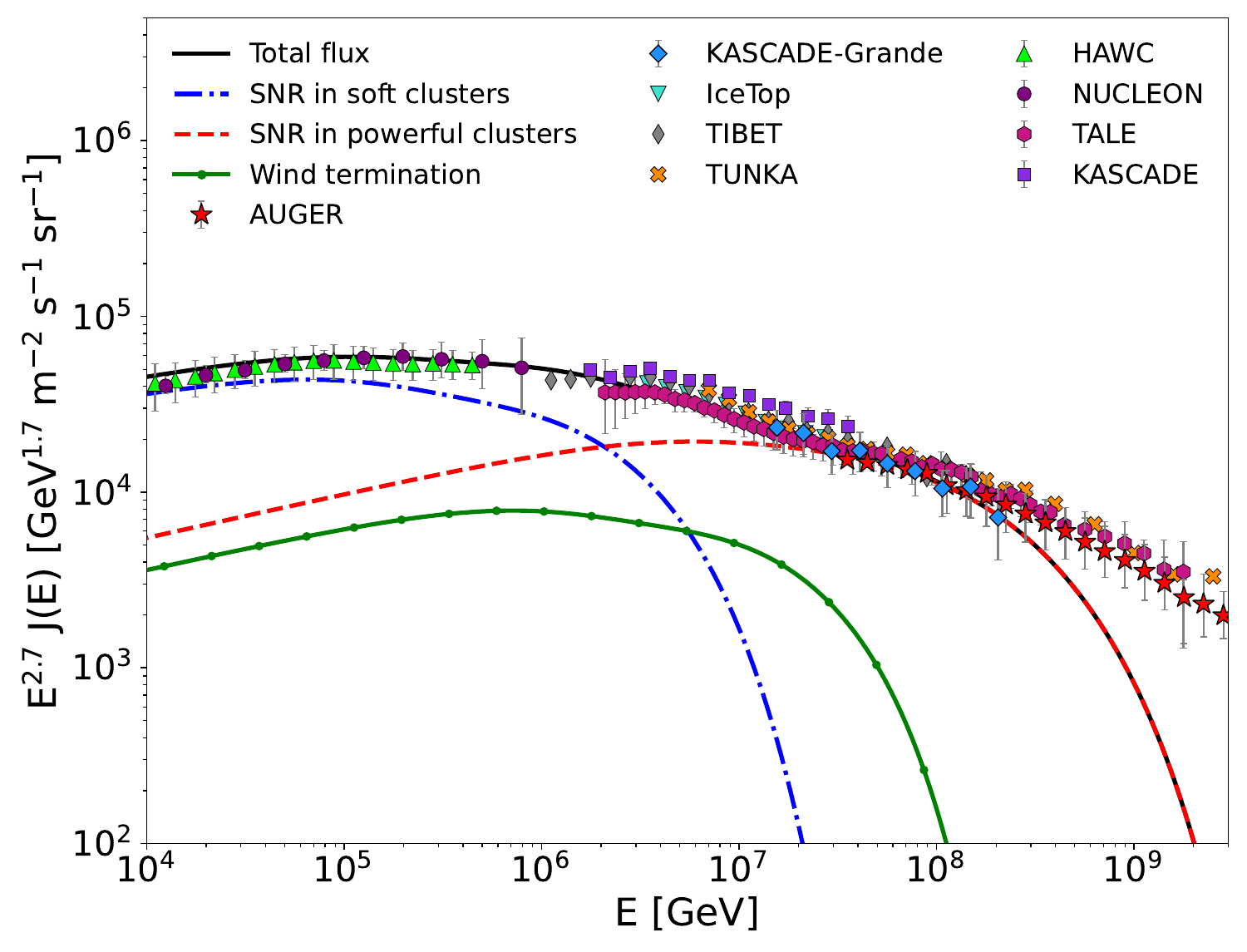}}  
    \subfloat[Corner plot for parameter constraints]{\includegraphics[angle=0,width=0.50\textwidth]{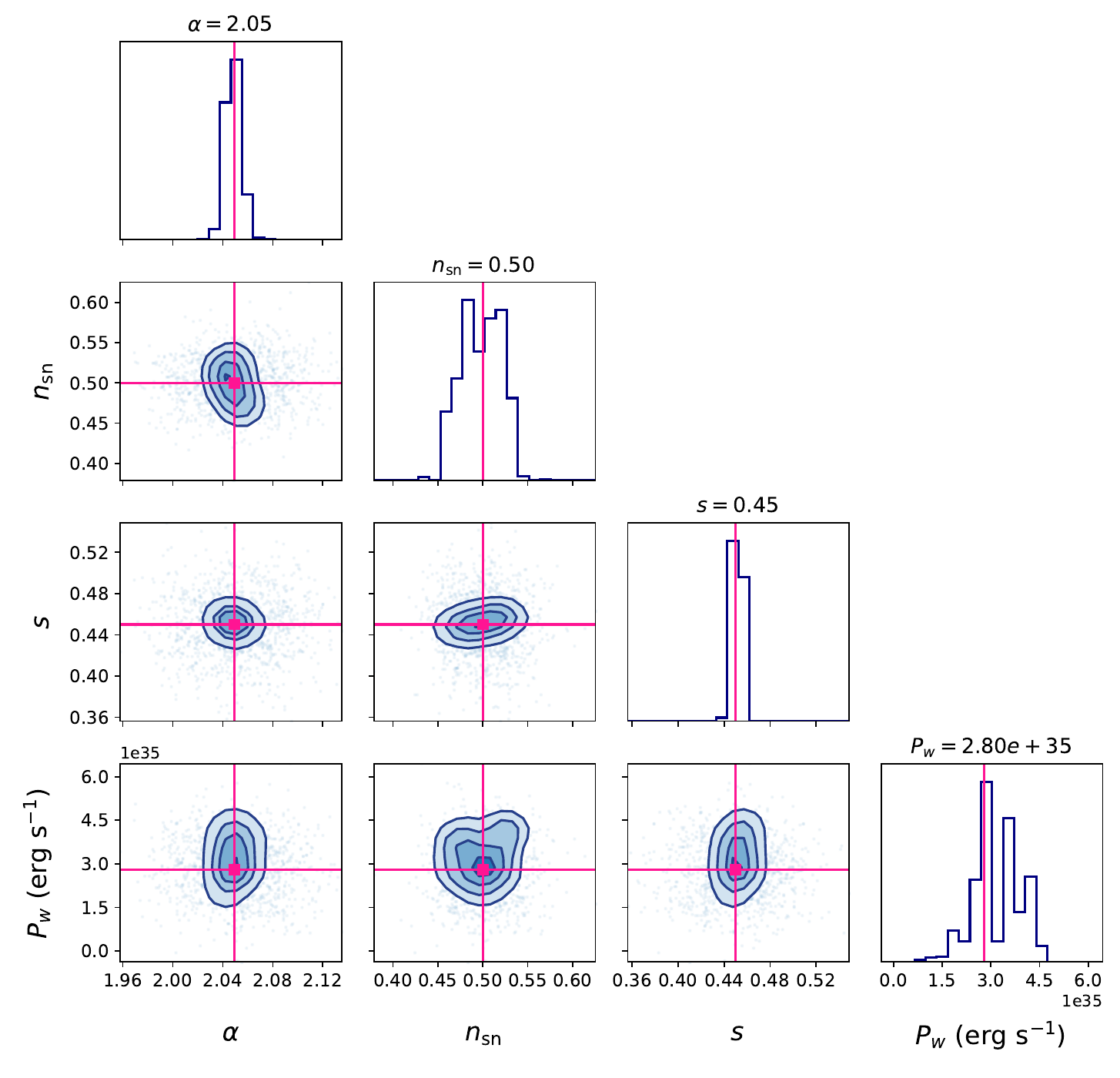}}
    \caption{(a) Modeled energy spectrum for accelerated nuclei up to charge $\mathrm{Z} = 40$, compared with recent experimental measurements from 
\protect\citet{2019arXiv190909073T,2017PhRvD..96l2001A,2021EPJC...81..966A,2019AdSpR..64.2546G,2017ICRC...35.1096M,2008ApJ...678.1165A,2020APh...11702406B}. (b) Best-fit model parameters obtained from fitting the spectrum.}
    \label{fig:spectra}
\end{figure*}

Massive stars produce strong stellar winds throughout their evolution, forming termination shocks~\citep{1980ApJ...237..236C,1983SSRv...36..173C}. In compact clusters, individual stellar shocks merge to form a collective wind that extends up to $\sim10$\,pc from the cluster center, where efficient particle acceleration occurs~\citep{2020MNRAS.493.3159G,2021MNRAS.504.6096M,2022MNRAS.515.2256V}. 
These collective winds can accelerate particles up to $E_{\rm wind}^{\rm max} \sim 1.0\,Z$\,PeV. 
This upper limit is determined by the balance between the acceleration time and the spatial scale of the wind termination shock. 
The presence of strong magnetic turbulence and high stellar-wind velocities enables efficient confinement of particles, allowing acceleration up to the PeV range~\citep{2021MNRAS.504.6096M,2024arXiv240316650M,2016A&A...591A..71M,2024NatAs...8..530P}.

The differential energy spectrum of accelerated nuclei with charge $Z$ is modeled as follows~\citep{2023MNRAS.519..136V}:
\begin{equation}
\begin{cases}
\phi_{\rm{soft}} =  \Gamma_{\rm{soft}} \, f_{\rm{Z}} \, \dfrac{E_{\rm{SN}}}{(p_0 c)^{2-\alpha}} \, E_p^{-\alpha} \, \exp\left(-\dfrac{E_p}{E_{\rm{soft}}^{\rm{max}}(\upsilon_5)}\right), \\[1.0em]
\phi_{\rm{pow}} =  \Gamma_{\rm{pow}} \, n_{\rm{sn}} \, f_{\rm{Z}} \, \dfrac{E_{\rm{SN}}}{2} \, E_p^{-\beta} \, \exp\left(-\dfrac{E_p}{E_{\rm{pow}}^{\rm{max}}(\upsilon_5)}\right), 
\end{cases}
\label{eq:spectrum}
\end{equation}
where $E_{\rm{soft}}^{\rm{max}}$ and $E_{\rm{pow}}^{\rm{max}}$, denote the maximum energies attained by particles accelerated in soft clusters and powerful clusters, respectively. The injection efficiency for each nuclear species is encapsulated in $f_{\rm{Z}}$, and $E_{\rm{SN}} \approx 10^{51}$ erg. The quantity $p_0$ sets the characteristic injection momentum, and $c$ is the speed of light. The normalization factors $\Gamma_{\rm{soft}}$ and $\Gamma_{\rm{pow}}$, determine the amplitude of each spectral component, and the spectral indices typically lie in the range $\alpha \approx 2.0$ – $2.3$ for soft clusters and $\beta \approx 2.0$ for powerful clusters. The term $n_{\rm{sn}}$ is 0.5 and denotes the fraction of supernovae contributing fast shocks within massive clusters (see Subsection~\ref{sec:2.3}). In our model, the particle escape time from the cluster is not explicitly calculated. We assume that it is short compared to the interaction time, allowing the spectrum of escaping particles to serve as a good approximation of the particle distribution inside the cluster.

The normalization constants are obtained by enforcing energy conservation across each component. Specifically, $\Gamma_{\rm{soft}}$ is defined such that the total energy injected by a supernova remnant is $f_{\rm Z} E_{\rm SN}$, while $\Gamma_{\rm{pow}}$ normalizes the powerful-cluster contribution to $\frac{1}{2} f_{\rm Z} E_{\rm SN}$, assuming that only part of the explosion energy is efficiently transferred to cosmic rays in environments with overlapping shocks.  

In each case, the normalization factor $\Gamma$ is determined by solving the following energy conservation condition
$\int_{E_{\min}}^{E_{\max}} E \cdot \phi(E)\, dE = E_{\rm tot}$,
where $E_{\rm tot}$ represents the total energy allocated to particle acceleration for the corresponding component (soft, powerful, or wind-driven), and the integration bounds define the energy range over which acceleration occurs. This ensures that the spectral amplitude is physically grounded and dimensionally consistent.
\begin{figure}
    \centering
    \includegraphics[width=1.0\linewidth]{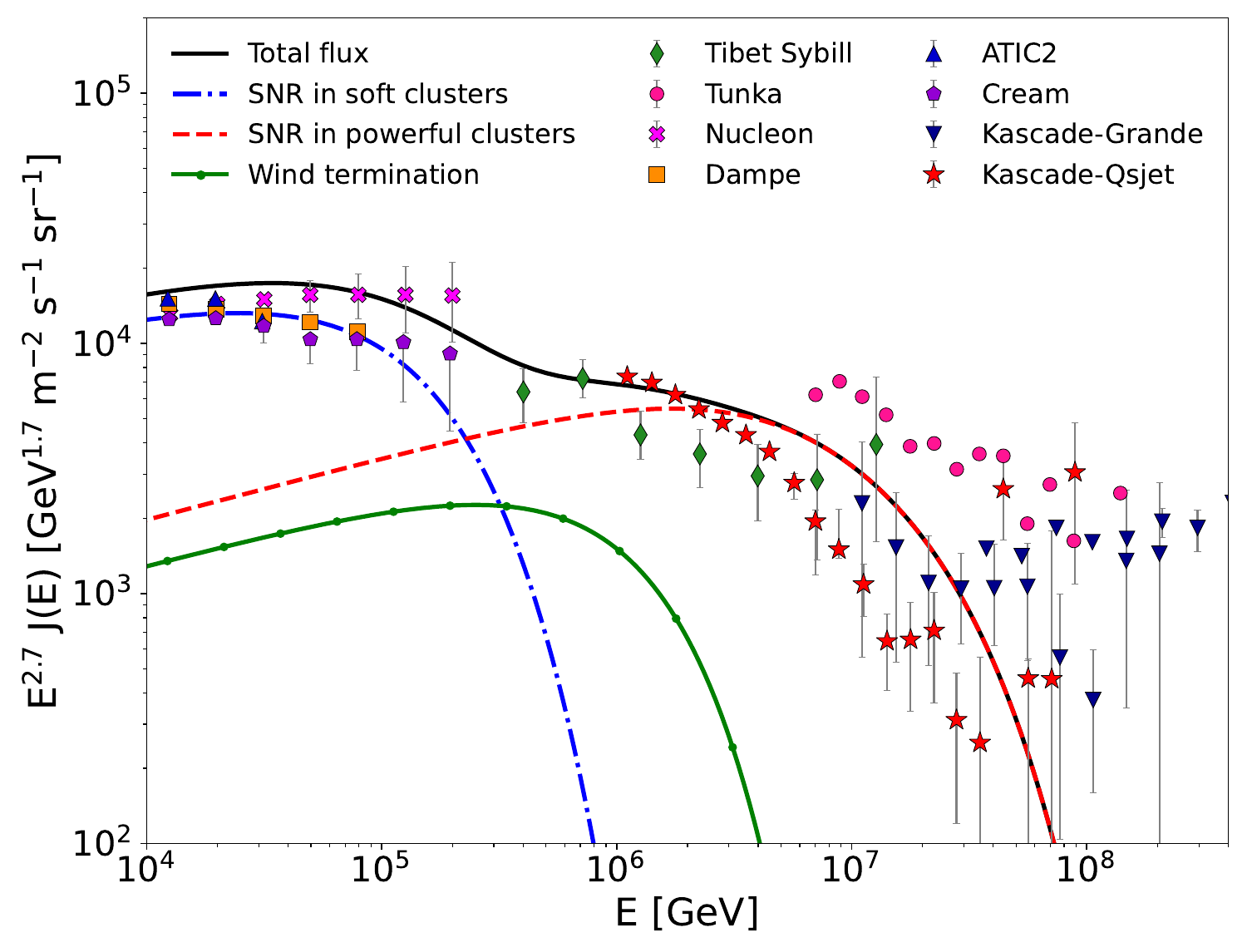}
    \caption{Proton spectrum predicted by the model (solid line) compared with experimental measurements from 
\protect\citet{2009BRASP..73..564P, 2019PhRvL.122r1102A, 2017ApJ...839....5Y, 2019SciA....5.3793A, 2005APh....24....1A, 2013APh....47...54A, 2019AdSpR..64.2546G, 2008ApJ...678.1165A}.}
    \label{fig:proton}
\end{figure}

The total Galactic cosmic-ray flux $\Phi$ from MSC, combining supernova and stellar wind contributions, is given by:
\begin{equation}
\begin{cases}
\Phi_{\rm{soft}} = f_{\rm{soft}} f_{\star} \int \phi_{\rm{soft}} \nu(\upsilon) d\upsilon, \\[1.0em]
\Phi_{\rm{pow}} = f_{\rm{pow}} f_{\star} \int \phi_{\rm{pow}} \nu(\upsilon) d\upsilon, \\[1.0em]
\Phi_{\rm{wind}} = \Gamma_{\rm{wind}} \, f_{\rm{Z}} \, E_p^{-2} \, \exp\left(-\dfrac{E_p}{E_{\rm{wind}}^{\rm{max}}}\right),
\end{cases}
\label{eq:fluxes}
\end{equation}
where $f_{\rm{pow}}$ and $f_{\rm{soft}}$ denote the fractional populations of powerful and soft clusters within $3\,\mathrm{kpc}$ (see Section~\ref{sec:2}), with numerical values listed in Table~\ref{tab:parameters}. he velocity distribution of shocks is taken as $\nu(\upsilon) \propto 1/\upsilon$, following the phenomenological prescription of~\citet{2023MNRAS.519..136V}. This functional form is not derived from first principles but rather reflects the decreasing probability of finding shocks at higher velocities, effectively modeling supernova shocks with a stepped tail (of order two high-velocity events per millennium). Such a choice has been tuned to reproduce the cosmic-ray spectrum measured by Auger around $1$\,EeV~\citep{2021EPJC...81..966A}. We constrain velocities to observed SN1987A limits: $\upsilon_{\rm min}$ from the late-phase deceleration~\citep{2018ApJ...867...65C} and $\upsilon_{\rm max}$ from its initial outburst, respectively (see Table \ref{tab:parameters}). 
Using these limits, the normalization of the distribution is fixed so that $\int_{\upsilon_{min}}^{\upsilon_{max}} \nu(\upsilon) d\upsilon \sim 0.02 \; \mathrm{yr}^{-1}$, in accordance with the Galactic core-collapse supernova rate inferred by~\citet{2023MNRAS.519..136V}.

The third term in Eq.~(\ref{eq:fluxes}), $\Phi_{\rm{wind}}$, represents the particle flux accelerated at the wind termination shocks of massive clusters.
Here, $E_{\rm{wind}}^{\rm{max}}$ corresponds to the maximum energy attained by particles accelerated in the wind termination region. For wind-driven acceleration, $\Gamma_{\rm wind}$ is normalized such that the integrated particle spectrum corresponds to the total power injected by stellar winds into accelerated particles, expressed as $f_{\rm Z}\,f_{\rm pow}\,f_{\star}\,N_{\rm Gal}\,P_{\rm w}$.
, where $N_{\rm Gal}$ is the Galactic number of massive stars, $P_{\rm w}$ is their mean mechanical wind luminosity in young clusters, and $f_{\star}$ quantifies the probability that a given massive star resides within a dense stellar environment~\citep{2023MNRAS.519..136V}.

The total propagated spectrum for nuclei of charge $Z$ can be written as
\begin{equation}
J(E,Z) = \left(\frac{c}{4\pi D_{\rm{gal}}}\right) \left(\frac{H^2}{V}\right)_{\rm{disk}} (\Phi_{\rm{soft}} + \Phi_{\rm{pow}} + \Phi_{\rm{wind}}),
\label{eq:totalflux}    
\end{equation}
where $H$ and $V$ are the Galactic halo’s scale height and volume, respectively. The Galactic diffusion coefficient is expressed as 
\[
D_{\rm gal} = D_0 \left(\frac{\mathcal{R}_p}{\mathcal{R}_0}\right)^{s},
\]
where \(s\) is the spectral index, \(D_0\) is the normalization constant, \(\mathcal{R}_p = E_p/Z\) is the particle rigidity, and \(\mathcal{R}_0 \sim 1~\mathrm{GV}\) is the reference rigidity corresponding to \(D_0 \sim \) 8 \(\times \;10^{28}~\mathrm{cm^2\,s^{-1}}\)~\citep{2024A&A...692A..20R}. 
This formulation is equivalent to the more common energy-based form for protons and remains valid up to the rigidity at which \(D/c \sim H\)~\citep{2023MNRAS.519..136V}.

The cosmic ray spectrum presented in Figures~\ref{fig:spectra}-(a) and~\ref{fig:proton} demonstrates three distinct components originating from MSC. The results shown in Figure~\ref{fig:spectra}-(a), where the solid black curve represents the sum of all components, were obtained from a parameter fit that minimizes the $\chi^2$ statistic, yielding a value of $\sim 7.068$ for the current observational data in the range $10^{4}$--$10^{8}$\,GeV. This relatively high $\chi^2$ value reflects the significant dispersion among the experimental measurements. The combined acceleration and injection components of the model, which include contributions from soft clusters, powerful clusters, and the wind termination shock, reproduce the observed cosmic-ray flux over the energy range $10^{4}$--$10^{8}$\,GeV.
This aligns with measurements of the Cygnus cocoon by \citet{2022AdSpR..70.2685B}, who reported a similar spectral slope and cutoff near 1~PeV, attributed to wind termination shocks in compact clusters. The normalization of this component is determined by the wind power, $P_{\mathrm{w}}$, which is treated in our model as a free parameter fitted to reproduce the observed cosmic-ray spectrum. The best-fit value, $P_{\mathrm{w}} = 2.8 \times 10^{35}\,\mathrm{erg.s^{-1}}$, is consistent with the average mechanical luminosity expected per massive star, including Wolf-Rayet (WR) stars of different types (WN, WC, WNb)~\citep{2007ARA&A..45..177C,2019A&A...621A..92S}. This fitted value represents the effective mean stellar-wind power within young clusters and ensures consistency with the Galactic cosmic-ray energy budget.

The powerful cluster component (red dashed line) dominates the spectrum between $4 \times 10^6$ and $5 \times 10^7$~GeV, exhibiting a harder spectral index ($\beta = 2.0$) and a maximum energy of $4\;Z$~PeV. These results are consistent with the predictions of \citet{2023MNRAS.519..136V} for young and dense clusters, where combined wind and supernova shocks accelerate particles to PeV energies. The adopted fraction of powerful clusters (11\% of the population) yields a flux normalization that agrees with the CR density measured near the 1 EeV \citet{2021EPJC...81..966A}. At lower energies ($10^4$ – $4 \times 10^6\,\mathrm{GeV}$), the soft cluster component (blue dashed line) exhibits a steeper spectrum, consistent with the CR flux measured in the Galactic plane by \citet{2012ApJ...750....3A}. The significant contribution of extended clusters in this band supports the scenario of isolated supernova shocks in evolved stellar populations \citep{2021MNRAS.504.6096M}. Furthermore, the transition between soft and powerful clusters around $\sim4\times10^5\,\mathrm{GeV}$ naturally reproduces the spectral knee at a few PeV as modeled by \citet{2016A&A...595A..33T}, resolving the discrepancy between single-source and multi-component models.

The corner plot in Figure~\ref{fig:spectra}-(b) quantifies the posteriors for our four free parameters, spectral index $\alpha$,  diffusion index $s$, and wind power $P_{w}$, with $\alpha$ showing the tightest constraint.
The fast-shock fraction parameter, $n_{\rm sn}$, is not a free parameter, it is included in the corner plot only to show its correlation with the other parameters. This reflects the relatively small statistical uncertainty on the spectral slope compared to the broader uncertainties in normalization, shock probability, and diffusion scaling~\citep{2013MNRAS.431..415B}. The best–fit model (black solid line) reproduces the spectral hardening and PeV-scale cutoff seen in the combined dataset, and the residuals between model and data remain below 15\% over the full energy range.

\begin{table}
\centering
\begin{tabular}{|c|c|}
\hline
\textbf{Parameter} & \textbf{Value} \\
\hline
 $E_{\rm{SN}}$ & $10^{51}$ [erg] \\
 $\left\{\begin{matrix} \upsilon_{\rm min} \\ 
 \upsilon_{\rm max} \end{matrix} \right. $ & $\begin{matrix} 3 000 \\ 
 30 000 \end{matrix}$ [km $/$ s] \\
\hline
$P_{\rm w}$ & $ 2.8 \times 10^{35}$ [erg $/$ s] \\
$N_{\rm gal}$ & $400 \times 10^{3}$ \\
\hline
$f_{*}$ & 0.80 \\
$f_{\rm{soft}}$ & 0.15\\
$f_{\rm{pow}}$ & 0.11 \\
$ n_{\rm{SN}} $ & 0.50 \\
\hline
$p_{0}$ & 0.01 [ GeV$/$c] \\
$f_{\rm Z}$ & \cite{2003APh....19..193H} [for 1 TeV] \\
$\alpha$ & 2.05 \\
$\beta$ & 2.00 \\
\hline
s & 0.45 \\
$D_0$ & 8 \(\times 10^{28}\) [cm$^2 / $ s ] \\
H & 3 [kpc] \\
V & 400 [kpc$^{3}$] \\
\hline
\end{tabular}
\caption{Parameters for the cosmic-ray propagation model. The nuclear injection efficiency ($f_{\rm Z}$) follows the prescription of~\protect\citet{2003APh....19..193H}, while velocity constraints are based on~\protect\citet{2023MNRAS.519..136V}. Remaining parameters are derived from observational constraints as discussed in Section~\ref{sec:2}.}
\label{tab:parameters}
\end{table}

The accelerated CR subsequently interacts with the interstellar medium, producing gamma rays and neutrinos primarily through pion production and decay~\citep{2022A&A...661A..72B,2019PhR...801....1A,2014CRPhy..15..357L}. The hadronic processes dominate, with neutral pions $\pi^0$ decaying to gamma rays via $\pi^0 \rightarrow 2\gamma$, while charged pions yield neutrinos through $\pi^\pm$ decay chains. This mechanism is particularly efficient in MSC environments due to their high gas density and CR flux~\citep{1994A&A...287..959D,2013Sci...342E...1I}. The energy spectrum of these secondary gamma rays provides critical constraints on CR propagation, as demonstrated by~\citet{Supanitsky_2013,Anjos_2014}. Recent IceCube measurements~\citep{2018Sci...361.1378I} further confirm the hadronic origin of this emission, while~\citet{2023ApJ...954....1C} have shown how the spatial distribution helps distinguish between leptonic and hadronic processes. In the next section, we use this particle injection model to calculate the photons produced by protons, following the analytical parameterizations for the inelastic $p$--$p$ collisions presented by~\citet{2014PhRvD..90l3014K}.

\section{Gamma Radiation Emission}\label{sec:4}

\begin{figure*}
   \centering
    \subfloat[Total Gamma]{\includegraphics[angle=0,width=0.505\textwidth]{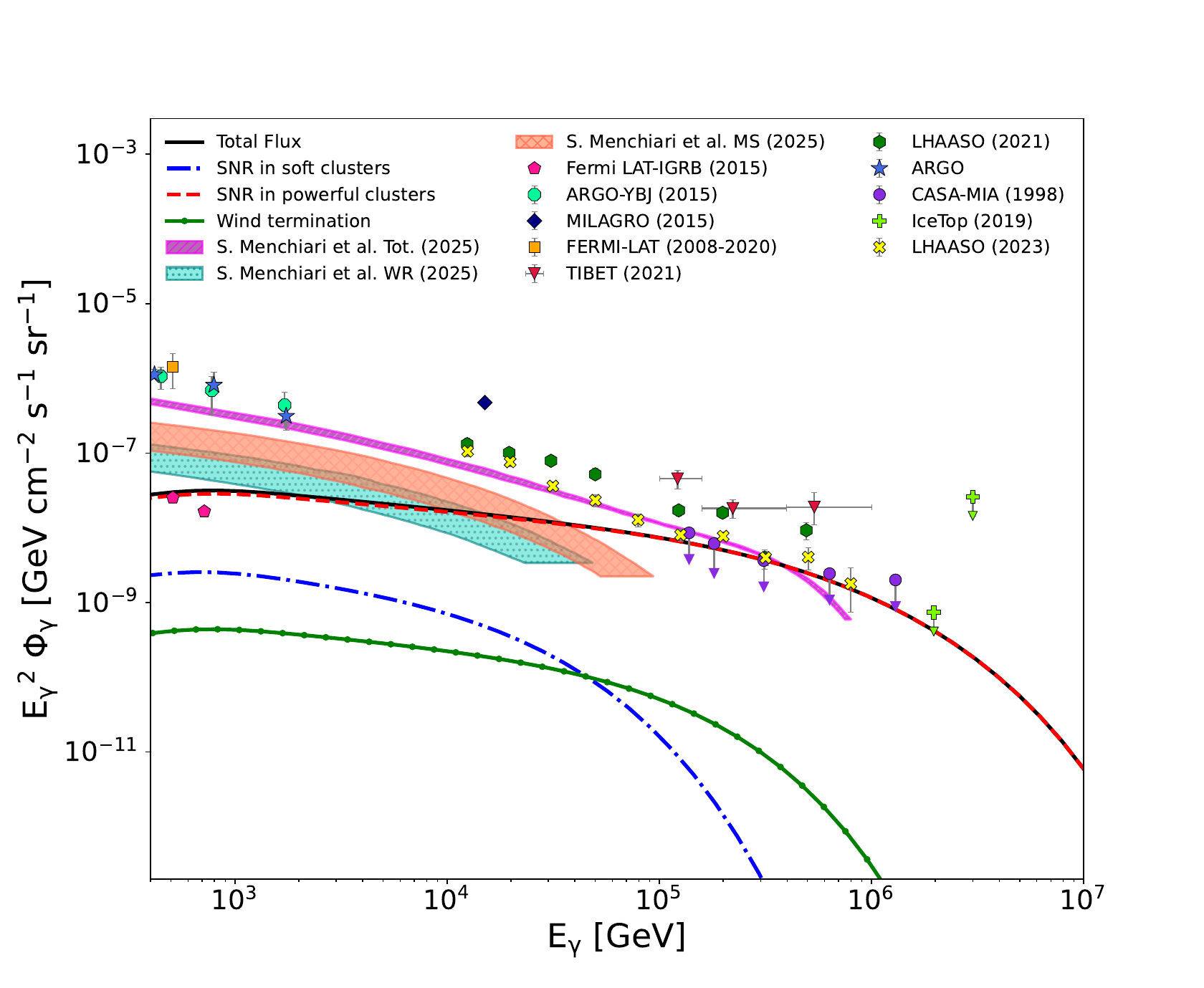}}
    \subfloat[Comparison with synthetic clusters and catalog + synthetic]{\includegraphics[angle=0,width=0.50\textwidth]{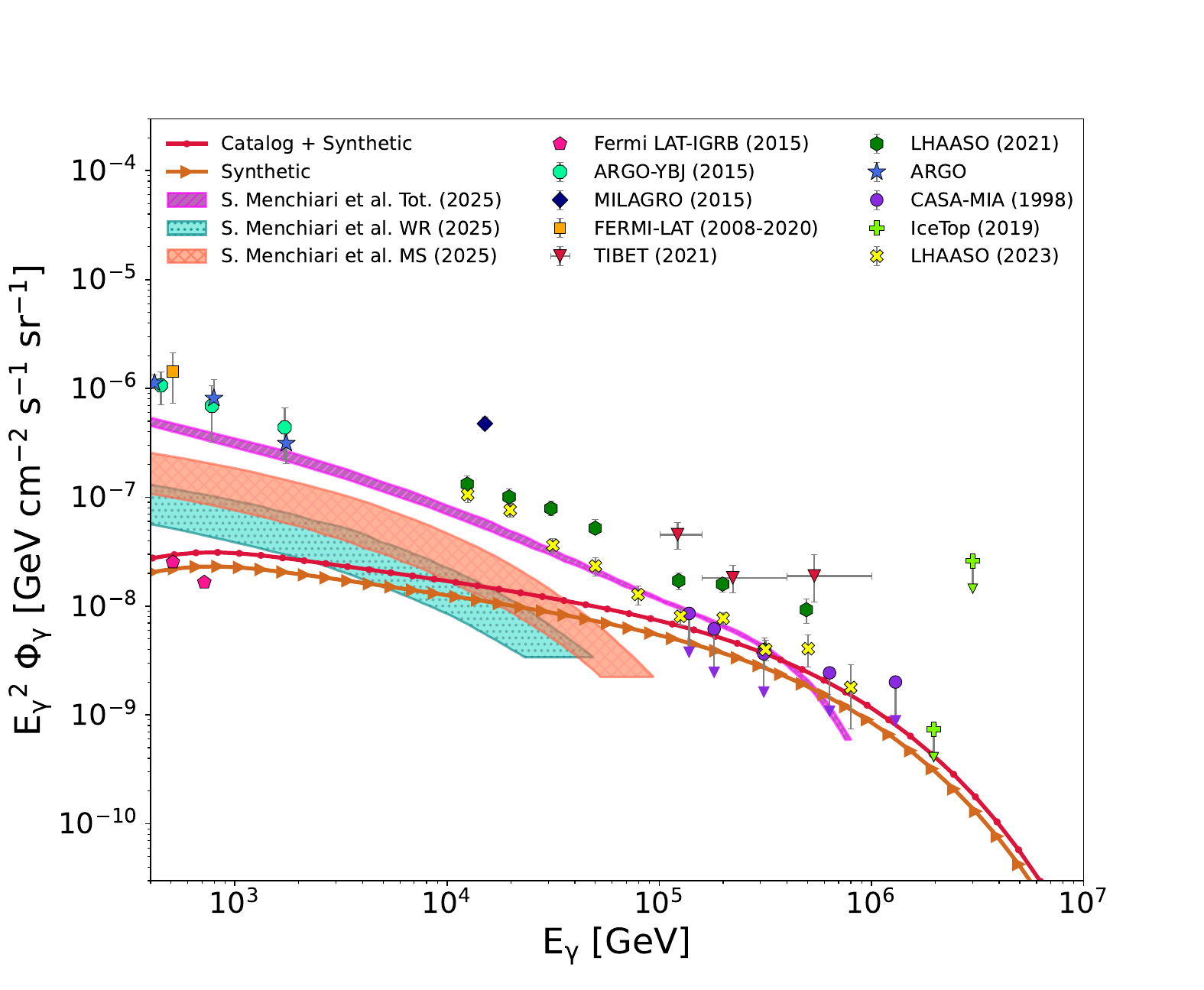}}  \caption{(a) Contribution to the gamma-ray emission compared with Galactic gamma-ray observations from \protect\citet{2015ApJ...806...20B,2017NPPP..291....9G,2021PhRvL.126n1101A,2009ApJ...692...61A,1998ApJ...493..175B,2017ApJ...849...67A,2023A&A...672A..58D,2015ApJ...799...86A,2023PhRvL.131o1001C}, as well as with the predictions from the models of~\protect\citet{2025A&A...695A.175M} adopting a Bohm diffusion coefficient.  (b) Simulated gamma-ray fluxes obtained from the combined contributions of soft and powerful clusters, comparing results derived from a fully synthetic catalog with those from real data supplemented by simulations. The cluster emission is normalized assuming a maximum particle energy corresponding to a shock velocity of $5000\,\mathrm{km\,s^{-1}}$. The data correspond to partial sky coverage by the individual instruments, whereas the model curves represent the total Galactic emission; no rescaling for sky coverage has been applied.}
    \label{fig:gamma}
\end{figure*}

Gamma-ray production can occur through hadronic and leptonic processes, including inverse Compton scattering, bremsstrahlung emission, and pion decay. 
It is well known that purely hadronic emissions are the most abundant source of gamma rays, representing one of the key phenomena in high-energy astrophysics. 
The main source of hadronic gamma rays comes from inelastic \(p\)--\(p\) collisions, which lead to the decay of light mesons~\citep{2019PhR...801....1A}. 
Additionally, heavier particles produced in these inelastic collisions decay rapidly, generating pions, which reinforces that the dominant source of gamma rays is the pion decay~\citep{2014PhRvD..90l3014K}. 
Given the high efficiency of gamma-ray production via pion decay, this work focuses on hadronic processes, specifically gamma rays generated by \(p\)--\(p\) collisions and proton interactions with interstellar medium nuclei. 

\begin{figure*}
   \centering
    \subfloat[]{\includegraphics[angle=0,width=0.50\textwidth]{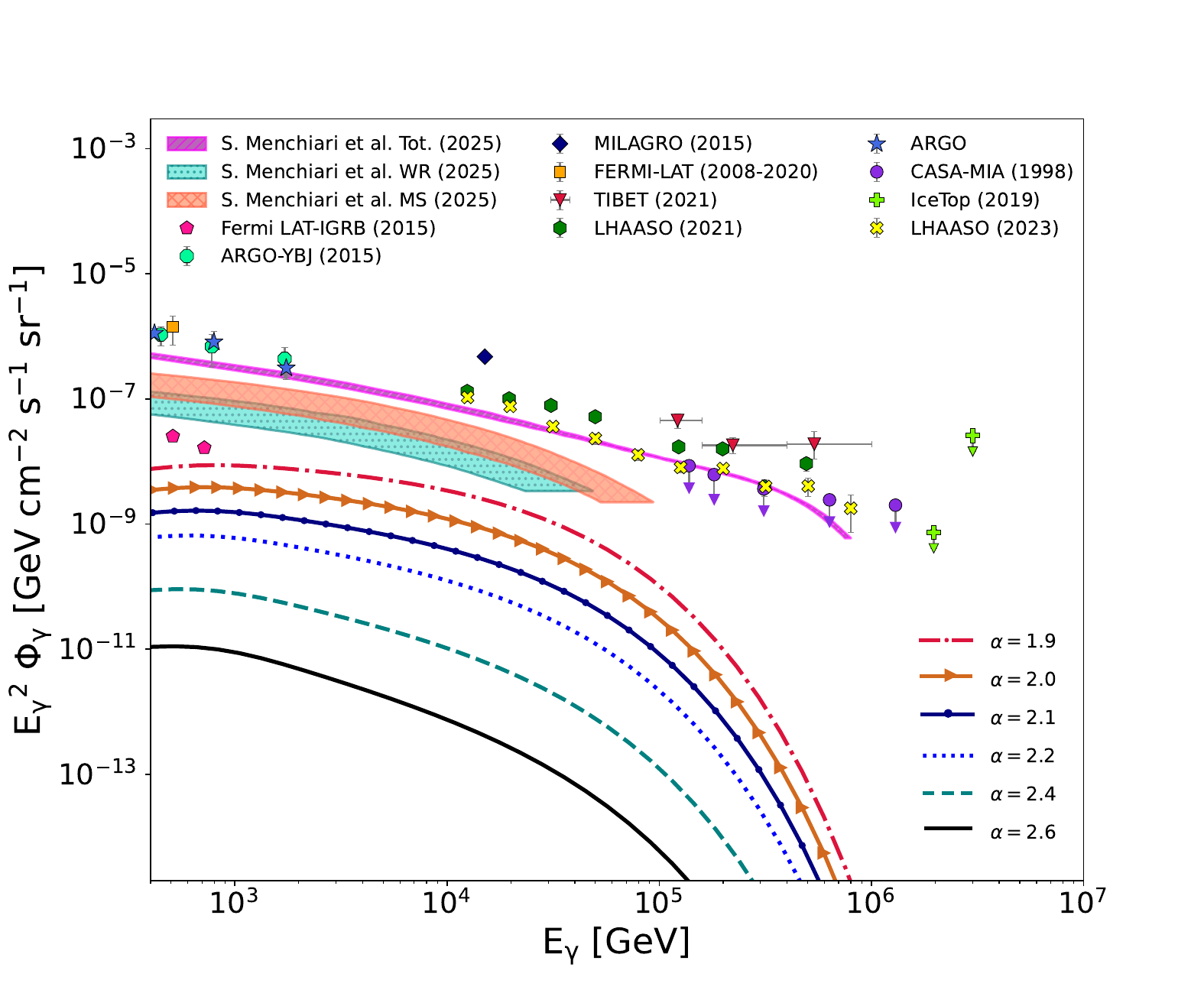}}
    \subfloat[]{\includegraphics[angle=0,width=0.50\textwidth]{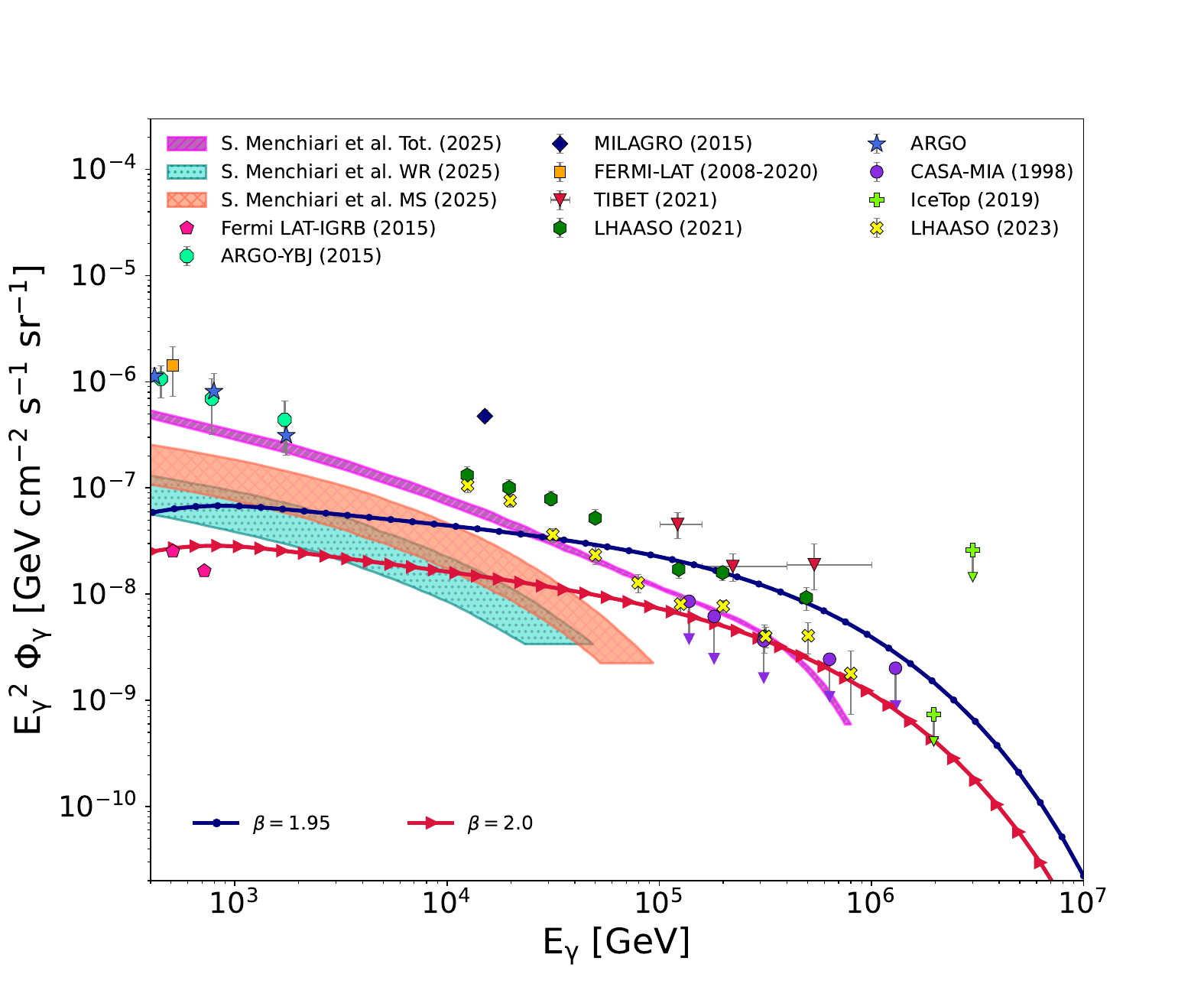}}  
    \caption{Gamma-ray emission from soft clusters (left) and powerful clusters (right) for different spectral indices. The resulting model contributions are compared with Galactic gamma-ray observations from 
\protect\citet{2015ApJ...806...20B}, \protect\citet{2017NPPP..291....9G}, \protect\citet{2021PhRvL.126n1101A}, 
\protect\citet{2009ApJ...692...61A}, \protect\citet{1998ApJ...493..175B}, \protect\citet{2017ApJ...849...67A}, 
\protect\citet{2023A&A...672A..58D}, \protect\citet{2015ApJ...799...86A}, and \protect\citet{2023PhRvL.131o1001C}, 
as well as with the predictions from the models of~\protect\citet{2025A&A...695A.175M} adopting a Bohm diffusion coefficient. The sky coverage and model curves follow what’s described in Figure \ref{fig:gamma}.}
    \label{fig:gamma_msc}
\end{figure*}

\begin{figure}
    \centering
    \includegraphics[width=1.00\linewidth]{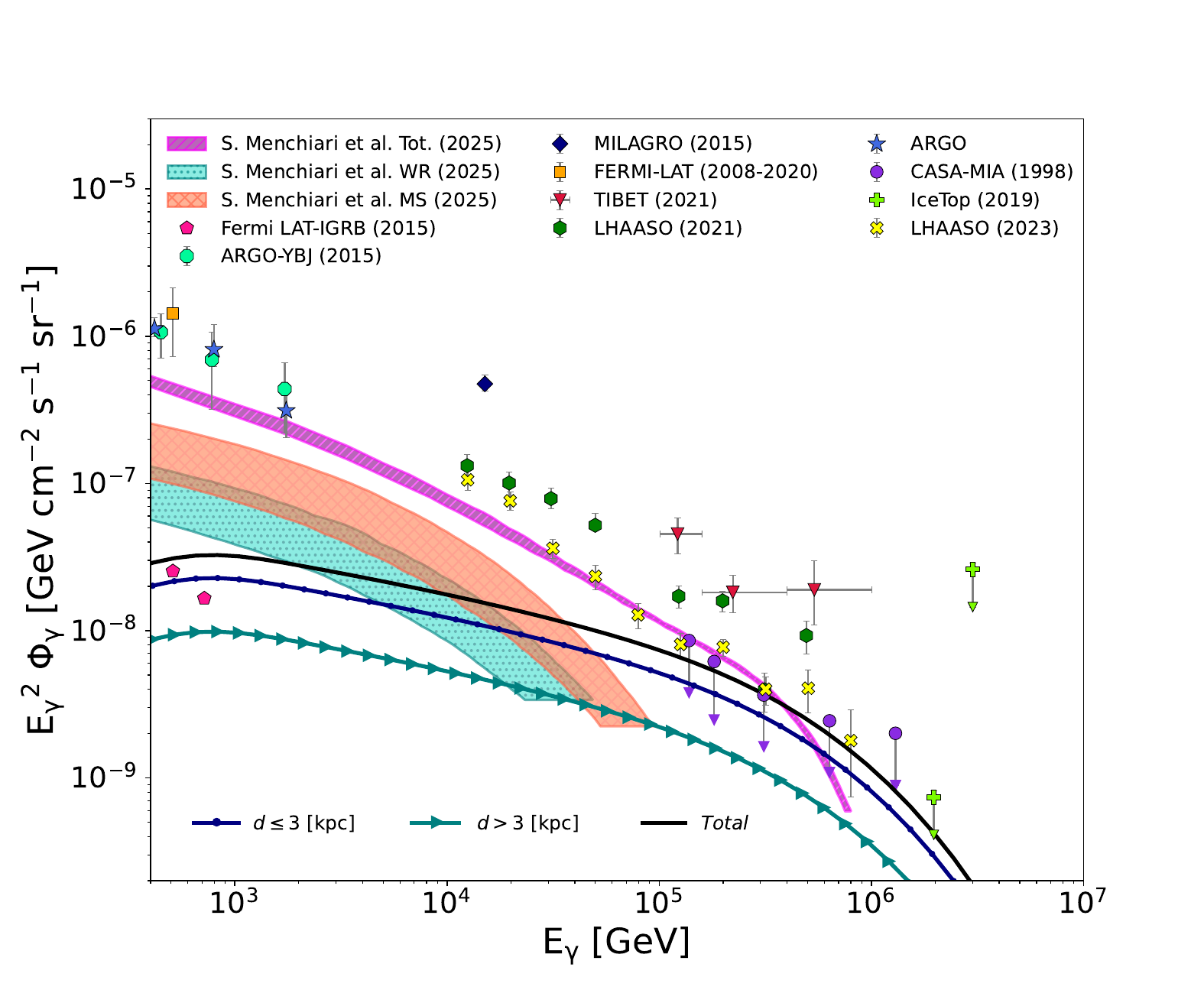}
    \caption{Gamma-ray fluxes combining the contributions from soft and powerful clusters within different distance limits. The figure shows how the total gamma-ray flux varies with the maximum source distance, in comparison with Galactic gamma-ray observations from 
\protect\citet{2015ApJ...806...20B,2017NPPP..291....9G,2021PhRvL.126n1101A,2009ApJ...692...61A,1998ApJ...493..175B,2017ApJ...849...67A,2023A&A...672A..58D,2015ApJ...799...86A,2023PhRvL.131o1001C}, 
as well as with the predictions from the models of~\protect\citet{2025A&A...695A.175M} adopting a Bohm diffusion coefficient. The sky coverage and model curves follow what’s described in Figure \ref{fig:gamma}.}
    \label{fig:cumulative}
\end{figure}

Currently, pion production is described by several models, including the isobar model~\citep{1957PhRv..105.1874L,1956PhRv..103..404Y}, which is based on baryon resonances, and the fireball model~\citep{1948PhRv...73..127L,1958PhRv..111.1699C,1950PThPh...5..570F,1952NW.....39...69H,1951PThPh...6..428U,1956NCim....3S..15B}, which treats the system as a hot pion gas using statistical methods. A combination of these approaches provides a spectral description of pion production~\citep{1968ApJ...151..881S}. Additional models fit accelerator data to the invariant differential cross section of pion production, from which the gamma-ray spectrum can be derived. In this work, we adopt the parameterization proposed by~\citet{2014PhRvD..90l3014K}, which provides a simple yet accurate expression for the gamma-ray differential cross section over a wide range of proton energies. This approach combines analytical modeling with experimental data and Monte Carlo simulations, ensuring consistency with accelerator measurements. To model the gamma-ray emission from MSC, we employ the population described in Section~\ref{sec:2}, which includes observational data from~\citet{2013A&A...558A..53K} for clusters within 3\,kpc and a complementary synthetic catalog for more distant sources. The total gamma-ray emission from Galactic MSC is then computed following the formalism of~\citet{2014PhRvD..90l3014K,2022MNRAS.512.1275V,2025A&A...695A.175M}:
\begin{equation}
    \Phi_{\gamma}(E_{\gamma}) = \frac{\eta_{\rm H}\; c}{4\pi d^2} \int \phi(E_p) \, \epsilon(E_p) \, \frac{d\sigma}{dE_{\gamma}}(E_p, E_{\gamma}) \, dE_p,
\end{equation}
where $d\sigma/dE_{\gamma}$ is the differential cross section for gamma-ray production from~\citet{2014PhRvD..90l3014K}, based on the SIBYLL model~\citep{1992PhRvD..46.5013E,2009PhRvD..80i4003A}; $E_p$ is the proton kinetic energy; and $\phi(E_p)$ is the injection spectrum (see Eq.~\ref{eq:spectrum}).  
The local gas density in the vicinity of clusters can differ significantly from the Galactic average, often being higher. For this reason, we adopt an average hydrogen density of $\eta_{\rm H} \sim 10\,\mathrm{cm^{-3}}$, which accounts for both the diffuse interstellar medium and denser regions within the Galaxy~\citep{2016SAAS...43...85K}, as well as the enhanced densities typically found around clusters~\citep{2022MNRAS.512.1275V,2025A&A...695A.175M,2025A&A...694A.244B,2024MNRAS.533..561B,2024icrc.confE.775C,2022A&A...667A..69S}. 
Here, $d$ is the distance from the Sun to the cluster, and $\epsilon(E_p) \simeq 1.15$ is the nuclear enhancement factor, calculated considering the most abundant nuclei in the interstellar medium: hydrogen, helium, carbon, and oxygen. We compute the gamma-ray emission individually for each cluster and sum their contributions to obtain the total flux from the MSC population. 

The gamma-ray emission from stellar wind termination shocks is estimated by assuming that the cosmic rays produced by the wind termination shock remain confined within the cluster's bubble throughout its lifetime. Therefore, this estimate provides an upper limit for the gamma-ray flux. However, despite this assumption, the final result shows that the wind contribution is negligible compared to that of the supernova remnants. Additionally, our model does not include an explicit treatment of particle escape from the cluster. Therefore, the resulting spectra should be considered a reasonable approximation up to energies around the TeV range. At higher energies, the predicted gamma-ray fluxes should be interpreted as upper limits. The total Galactic gamma-ray flux from MSC is then obtained by summing the contributions of all individual clusters. In Figure ~\ref{fig:gamma}, the model curves represent the integrated emission from the entire Galactic MSC population. Although the experimental data correspond to specific sky regions, the dominance of nearby clusters, particularly below $\sim10^{6}$\,GeV, makes the spectral comparison meaningful.

Figure~\ref{fig:gamma} presents the modeled gamma-ray emission from MSC along with the theoretical model by~\citet{2025A&A...695A.175M}, which includes both the stellar-cluster and diffuse Galactic cosmic-ray components. Panel~\ref{fig:gamma}-(a) shows the differential gamma-ray flux resulting from inelastic $p$–$p$ collisions, calculated using the parameterization of~\citet{2014PhRvD..90l3014K}. This model incorporates gamma-ray contributions from three distinct MSC components: soft clusters, powerful clusters, and wind termination shocks. The black curve represents the total predicted flux from all components combined. The agreement at high energies underscores the relevance of pion decay as a dominant gamma-ray production mechanism in MSC.  Panel~\ref{fig:gamma}-(b) compares gamma-ray flux predictions derived from a fully synthetic MSC population with those obtained from a hybrid catalog combining observed clusters \citep{2013A&A...558A..53K} within 3 kpc and synthetic clusters beyond that distance. Both scenarios include contributions from soft and powerful clusters, with emission normalized using a maximum particle energy corresponding to a shock speed of $5000\,\mathrm{km\,s^{-1}}$. 

Figure~\ref{fig:gamma_msc} illustrates aspects of the gamma-ray emission originating from MSC under different physical assumptions.
Figure~\ref{fig:gamma_msc}-(a) explores the impact of varying the spectral index on the gamma-ray emission from soft clusters. The curves correspond to different values of the spectral index $\alpha$, ranging from 1.9 to 2.6, and clearly demonstrate how the slope and normalization of the resulting gamma-ray spectrum change with this parameter. Softer spectra (larger $\alpha$) yield steeper gamma-ray curves, reducing the high-energy component of the emission. Since soft clusters are associated with older, less dense environments where particle acceleration is less efficient, this spectral variation introduces only a moderate deviation in the model predictions at higher energies. Nonetheless, such variation could be significant when comparing to models like~\citet{2025A&A...695A.175M}, which emphasize detailed spectral modeling of young clusters. In addition, Figure~\ref{fig:gamma_msc}-(b) focuses on powerful clusters and compares gamma-ray spectra computed using two close spectral indices: $\alpha = 1.95$ and $\alpha = 2.00$. Although the numerical difference between the indices is small, the resulting gamma-ray fluxes diverge significantly at high energies. This strong sensitivity is a natural consequence of power-law behavior: harder spectra (lower $\alpha$) maintain a relatively larger population of high-energy protons, which in combination with the energy dependent nature of the $p$--$p$ cross section, results in a substantially amplified gamma-ray output at TeV–PeV energies. In contrast, the slightly softer index of $\alpha = 2.00$ leads to a steeper decline, suppressing the contribution from the most energetic particles. This behavior shows that even minor variations in the injection index can produce differences in predicted fluxes at the upper end of the spectrum, emphasizing the importance of accurate spectral constraints when modeling the gamma-ray emission from powerful clusters.

Figure~\ref{fig:cumulative} shows the cumulative gamma-ray flux from MSC, emphasizing the role of spatial distribution in shaping the observed emission. The curves separate the contributions from clusters located within 3\,kpc from the Sun, beyond 3\,kpc, and the combined total (black). The results indicate that nearby clusters can dominate the emission up to $\sim10^{6}$\,GeV, depending on their spatial distribution, the adopted distance threshold used to define the local population, and the completeness of the available cluster catalogs. The predicted spectrum is compared to gamma-ray data from several experiments, as well as with the model presented by~\citet{2025A&A...695A.175M}, showing good agreement in both normalization and spectral shape. Additionally, Figure~\ref{fig:cumulative} shows the consistency between the two approaches at high energies, confirming that nearby MSC dominate the gamma-ray flux, while contributions from clusters beyond 3 kpc is only $\sim 1/3$ of the integrated emission.

\section{Conclusions}\label{sec:5}

In this work, we investigated the role of MSC as sources of high-energy cosmic rays and gamma-ray emission in the Milky Way. By modeling proton injection and acceleration processes associated with both supernova explosions and collective stellar winds, we examined how different classes of clusters, classified as soft and powerful systems, contribute to the observed cosmic-ray and gamma-ray spectra, following the framework developed by \citet{2023MNRAS.519..136V}. Our approach combines observational data from the \citet{2013A&A...558A..53K} catalog, which is complete within approximately 3\,kpc of the Sun, with a synthetic population of clusters distributed across the Galactic disk to compensate for incompleteness at larger distances. This hybrid method enables a more representative treatment of the Galactic MSC population.

The particle spectrum was constructed by incorporating three principal components: acceleration from isolated supernova remnants in soft clusters, the combined effects of supernova and wind-driven shocks in powerful clusters, and continuous acceleration at wind termination shocks. Model parameters were calibrated to reproduce observational signatures, including the spectral slope, the cutoff at PeV energies, and the normalization of the total cosmic-ray flux, in agreement with other recent studies such as \citet{2023MNRAS.519..136V}, \citet{2022A&A...661A..72B}, \citet{2021MNRAS.504.6096M}, and \citet{2020MNRAS.493.3159G}. The resulting particle distributions were then used to compute the gamma-ray flux, assuming hadronic \(p\)--\(p\) interactions as the dominant emission mechanism.

The predicted gamma-ray spectrum shows reasonable agreement with measurements reported by high-energy experiments, particularly in the range $\sim10^{5}$–$10^{6}$\,GeV (see Figure~\ref{fig:gamma}). This agreement supports the interpretation that MSC, particularly young and compact systems with strong stellar winds, are important contributors to the Galactic gamma-ray emission. These powerful clusters provide favorable conditions for particle acceleration to PeV energies, naturally explaining features of the observed spectrum, including the hardening near the so-called ``knee'' in the cosmic-ray distribution. Our results also indicate that clusters within a radius of 3\,kpc dominate the gamma-ray flux up to $\sim 10^6$\,GeV, consistent with the spatial completeness of existing catalogs in the solar neighborhood. It should be noted that the model predictions correspond to an all-sky integrated gamma-ray flux, whereas the observational data used for comparison were obtained from specific regions of the sky. 
Consequently, the agreement discussed here is intended to be qualitative, since variations in sky coverage and instrumental exposure among the different experiments can influence both the normalization and the spectral shape of the measured flux.

While the model reproduces the overall spectral trend, it does not attempt to describe the diffuse Galactic gamma-ray background. 
Differences that appear at intermediate energies or for regions beyond 3\,kpc likely reflect this limitation, as the diffuse emission includes contributions from older or lower-mass clusters and other astrophysical sources not considered here.
Furthermore, the model’s sensitivity to parameters such as wind luminosity and injection spectral index highlights the importance of obtaining more precise observational constraints on cluster properties and the underlying physics of particle acceleration.

In conclusion, our study reinforces that MSC, especially the youngest and most compact among them, play a central role in shaping the high-energy emission of our Galaxy. The methodology adopted here provides a framework for interpreting gamma-ray observations in terms of both resolved and unresolved cluster populations. Future work should prioritize advancing stellar cluster population models, improving the physical treatment of wind-driven shocks and their role in particle acceleration, and integrating multi-messenger constraints from neutrinos, cosmic rays, and non-thermal emission. These combined advances will bridge gaps between theory and observations of high-energy processes in compact stellar systems.

\section*{Acknowledgements}
We sincerely thank the referees for their thoughtful feedback and valuable suggestions, which have greatly enhanced the clarity and scientific rigor of this work. We are particularly grateful to Brian Reville and Thibault Vieu for their discussions and contributions throughout this project. We also thank Rubens Costa and Rubens Machado for their stimulating conversations on particle propagation and stellar cluster dynamics. L.N.P acknowledges financial support from the Coordenação de Aperfeiçoamento de Pessoal de Nível Superior – Brasil (CAPES) – Finance Code 001. L.N.P. and R.C.A. acknowledge the support of the NAPI “Fenômenos Extremos do Universo” of Fundação de Apoio à Ciência, Tecnologia e Inovação do Paraná. R.C.A. research is supported by CAPES/Alexander von Humboldt Program (88881.800216/2022-01), CNPq (310448/2021-2) and (4000045/2023-0), Araucária Foundation (698/2022) and (721/2022) and FAPESP (2021/01089-1). R.C.A. gratefully acknowledges the Max Planck Institute for Nuclear Physics for their warm hospitality and support during her visit, which provided a conducive environment for fruitful discussions and collaborations. R.C.A. also acknowledges the support of L’Oreal Brazil, with the partnership of ABC and UNESCO in Brazil. The authors acknowledge the AWS Cloud Credit/CNPq and the National Laboratory for Scientific Computing (LNCC/MCTI, Brazil) for providing HPC resources of the SDumont supercomputer, which have contributed to the research results reported in this paper. URL: https://sdumont.lncc.br.

%%%%%%%%%%%%%%%%%%%%%%%%%%%%%%%%%%%%%%%%%%%%%%%%%%
\section*{Data Availability}

No new data were generated or analysed in support of this research.

%%%%%%%%%%%%%%%%%%%% REFERENCES %%%%%%%%%%%%%%%%%%

% The best way to enter references is to use BibTeX:

\bibliographystyle{mnras}
\bibliography{example} % if your bibtex file is called example.bib

% Alternatively you could enter them by hand, like this:
% This method is tedious and prone to error if you have lots of references
%\begin{thebibliography}{99}
%\bibitem[\protect\citeauthoryear{Author}{2012}]{Author2012}
%Author A.~N., 2013, Journal of Improbable Astronomy, 1, 1
%\bibitem[\protect\citeauthoryear{Others}{2013}]{Others2013}
%Others S., 2012, Journal of Interesting Stuff, 17, 198
%\end{thebibliography}

%%%%%%%%%%%%%%%%%%%%%%%%%%%%%%%%%%%%%%%%%%%%%%%%%%

%%%%%%%%%%%%%%%%% APPENDICES %%%%%%%%%%%%%%%%%%%%%

%\appendix

%\section{Some extra material}

%f you want to present additional material which would interrupt the flow of the main paper,
%it can be placed in an Appendix which appears after the list of references.

%%%%%%%%%%%%%%%%%%%%%%%%%%%%%%%%%%%%%%%%%%%%%%%%%%

% Don't change these lines
\bsp	% typesetting comment
\label{lastpage}
\end{document}